\documentclass[aps,pre,twocolumn,amsfonts,amssymb,showpacs]{revtex4}
\usepackage{amsmath}
\usepackage{amssymb}

\begin{document}

\title{Specific features of the effect of time dependent field on subdiffusing particles.
The stochastic Liouville equation approach.}
\author{A. I. Shushin}
\affiliation{Institute of Chemical Physics, Russian Academy of
Sciences, 117977, GSP-1, Kosygin str. 4, Moscow, Russia}

\begin{abstract}
We analyze the effect of time dependent external field on
non-Markovian migration described by the continuous time random walk
(CTRW) approach. The rigorous method of treating the problem is
proposed which is based on the Markovian representations of the CTRW
approach and field modulation. The method is applied to the case of
subdiffusive migration in which the exact formulas for the first and
second moments of spatial distribution are derived. For oscillating
external field they predict unusual dependence of the first moment
on oscillation phase and anomalous field dependent contribution to
the dispersion. Similar formulas are also derived fluctuating field.
\end{abstract}

 \maketitle

\bigskip

\section{Introduction.}

Brownian motion in external time-dependent field is the important
stage of many physical and chemical processes which often strongly
affect their kinetics \cite{Han,Rei}. Last years close attention is
given to the anomalous (subddifusive) jump-like motion typical for
disordered systems \cite{Sch,Met1} and, in particular, to the effect
of time-dependent field on this type of migration \cite{Sok, Han1}.
Usually the motion anomaly is assumed to be a manifestation of the
long memory in the kinetics of jumps. In such a case the serious
difficulty in theoretical treatment of time-dependent field effects
occurs because of subtle interplay of field and anomalous memory
effects which should be properly described.

Subdiffusive processes in time-independent potential $V (x)$ are
traditionally described by the fractional Smoluchowski equation
(FSE) for the probability distribution function (PDF) $\rho (x,t)$
\cite{Met1}
\begin{equation} \label{field1}
\dot \rho= -\, {}_0^{}\!D_t^{1-\alpha}\hat {\cal L}_{\alpha} \rho,
\end{equation}
where ${}_0^{}\! D_t^{1-\alpha}$ is the Riemann-Liouville fractional
derivative defined by
\begin{equation}
{}_0^{}\! D_t^{1-\alpha} \psi = \frac{1}{\Gamma( \alpha)}
\frac{\partial}{\partial t} \int_0^t d\,t_1\,\frac{\psi (t_1)}{ (t -
t_1)^{1-\alpha}}
\end{equation}
and
\begin{equation} \label{field3}
\hat {\cal L}_{\alpha} = -D_{\alpha} \nabla_{x} [\nabla_{x} -  F(x)]
\end{equation}
is the Smoluchowski operator, in which $D_{\alpha}$ is a
subdiffusion constant, $\nabla_{x} \equiv
\partial/\partial x$, and $F(x) = -\nabla_x V(x)/(k_B T)$ is a
force. The FSE (\ref{field1}) can be derived within the continuous
time random walk (CTRW) approach \cite{Met1} assuming the long time
tailed behavior of the waiting time distribution $W(t)$ for
CTRW-jumps: $W(t) \sim 1/t^{1+\alpha} \, (\alpha < 1)$.

In the case of time-dependent field $F(x,t)$ [i.e. time-dependent
${\cal L}_{\alpha}(t)$], however, no analogs of the FSE are
rigorously derived as yet. The main difficulty is in correct
treatment of the effect of ${\cal L}_{\alpha}(t)$-evolution during
the time of waiting for jumps. Only approximate variants of these
FSEs have been proposed so far \cite{Sok, Han1}.

In this work within the CTRW approach we derive the exact FSE
(describing the influence of time-dependent field)  with the use of
recently proposed Markovian representation of the CTRW and the
non-Markovian stochastic Liouville equation (SLE) \cite{Shu1}. The
solutions of this FSE for different time dependences of the force
$F(t)$, for simplicity, assumed to be independent of $x$, are
proposed and discussed in detail.

\section{Markovian SLE.}

Here we will present the method of treating the effect of
time-dependent field $F(x,t)$ on CTRW-like processes by reduction of
the problem to solving the SLE with time-independent operators.

To clarify the method we first consider the Markovian (normal
diffusion) case: $\alpha = 1$, in which the evolution of the system
is described by the Smoluchowski equation
\begin{equation} \label{field4}
\dot \rho= -\hat {\cal L}_1(t) \rho = D_1 \nabla_{x} [\nabla_{x}\rho
- F(x,t)\rho].
\end{equation}

The method is based on the representation of the time dependence of
$F(x,t)$ in terms of the dependence on some Markovian (in general,
stochastic) variable $z(t)$: $F(x,t) \equiv F(x,z(t))$, whose
evolution is described the PDF $\sigma (z,t)$ satisfying the
Markovian equation
\begin{equation} \label{field5}
\dot \sigma = -\hat {L} \sigma, \;\;\mbox{with}\;\;\sigma
(z,0)=\sigma_i (z),
\end{equation}
in which $\hat L$ is the linear operator in $\{z\}$-space and $\int
\! dz\,\sigma_i (z) = 1$. For brevity, formulas are written assuming
that $\{{\bf z}\}$-space is one dimensional, though they are,
evidently, valid for any dimensionality of $\{{\bf z}\}$-space. The
corresponding examples will be discussed below. In addition, in what
follows we will restrict ourselves to the simple model of
$x$-independent force:
\begin{equation} \label{field6}
F(x,t) \equiv F(x,z(t)) = F_0 z(t).
\end{equation}

In this representation the kinetics of the process described by eq.
(\ref{field4}) is determined by the average evolution operator which
in the space $ \{x\otimes z\}$ is given by formula
\begin{equation} \label{field7}
U(x,z;x_i,z_i|t) = \langle x,z|\langle Te^{-\int_0^t d\tau \hat
{\cal L}_{1}(\tau)} \rangle|x_i,z_i\rangle,
\end{equation}
where the average (denoted as $\langle \dots \rangle$) is taken over
trajectories of the stochastic Markovian process in $\{z\}$-space
with fixed initial ($z_i$) and final ($z$) coordinates. In
particular, the PDF of interest, $\overline{\rho}_F (x,x_i|t)$
averaged over $F(t)$-fluctuations, can be calculated as
\begin{equation} \label{field8}
\overline{\rho}_F (x,x_i|t) = \!\int \!dz\!\int \!dz_i \,
U(x,z;x_i,z_i|t)\sigma_i(z_i).
\end{equation}

The important point of the proposed representation consists in the
fact that for Markovian processes in $\{x\otimes {\bf z}\}$-space
the operator $\hat U$ is satisfies the Markovian SLE with
time-independent operators \cite{Kubo}
\begin{equation} \label{field9}
\dot {\hat U} = -(\hat {\cal L}_1 + \hat L) \hat U \;\;
\end{equation}
with  $\,U(x,z;x_i,z_i|0) = \delta({x\!-\!x_i})\delta({z\!-\!z_i})$.

Thus we have reduced the problem to solving the SLE (\ref{field9})
with time-independent operators, though at the cost of the extension
of the space of the process, which describes the evolution of the
system.

Noteworthy is that the representation (\ref{field7})-(\ref{field9})
is valid not only for stochastic functions $z(t)$ but also for
dynamical ones, which are known to be Markovian as well. For
example, in the model of harmonically oscillating force:
\begin{equation} \label{field10}
z(t) = z_c(t)= z_0 \sin (\omega t + \varphi),
\end{equation}
the dependence $z_c (t)$ can be considered as a coordinate part of
the trajectory of dynamical motion (in the harmonic potential),
described by the operator
\begin{equation} \label{field11}
\hat L = -(v\nabla_z + \omega^2 z \nabla_{v_z}),
\end{equation}
in the phase space $\{{\bf z}\} = (z,v_z)$ ($v_z =\dot z$ is the
velocity) with $\sigma_i ({\bf z})=\delta(z\!-\!z_0{\sin
\varphi})\delta(v_z\!\! -\! z_0\cos\varphi)$. Evidently, the case of
$z(t)$ represented as a linear combination of, say, $N$ oscillating
functions $z_j(t) = z_{_{0_j}} \sin (\omega_j t + \varphi_j)$: $z(t)
= \sum_0^{N} z_j(t)$, can be modeled by coupling to $N$ harmonic
coordinates ${\bf z}_N = (z_1, z_2,\dots, z_N)$.

\section{Non-Markovian CTRW.}

\subsection{Markovian representation.}

The main goal of this work is the analysis of the effect of
time-dependent field on CTRW-type (subdiffusive) migration.

In the CTRW approach the stochastic motion in $\{x\}$-space is
treated as a set of jumps with jump statistics described by the
waiting time distribution $W(t)$ \cite{Sch,Met1}. For
time-independent driving force the non-Markovian equation for the
PDF $\rho (x,t)$ in is conventionally derived by summing up the
contributions of all sets of jumps. In terms of the Laplace
transform $R (\epsilon) = \int_0^{\infty} \!dt \, \rho
(t)e^{-\epsilon t}$, this equation is written as \cite{Sch,Met1}
\begin{equation} \label{nmark1}
\epsilon R (\epsilon) = \rho_i - M(\epsilon) \hat {\cal L}_{\alpha}
R (\epsilon).
\end{equation}
In this equation $\rho_i(x)$ is the initial PDF and
\begin{equation} \label{nmark2}
{M}(\epsilon) = [1-\widetilde W(\epsilon)]/[\epsilon \widetilde
W(\epsilon)],
\end{equation}
where $\widetilde {W} (\epsilon) = \int_0^{\infty} \!dt \,
W(t)e^{-\epsilon t}$. Note that in the case of subdiffusion, when $M
(\epsilon) = \epsilon^{1-\alpha}$, eq. (\ref{nmark1}) reduces to the
Laplace transform variant of the FSE (\ref{field1}).

CTRW-type processes can conveniently be analyzed within the
Markovian representation \cite{Shu1} in which these processes are
assumed to result from jump-like $\hat {\cal L}_1(t)$-fluctuations
determined by the dependence  of $\hat {\cal L}_1 (y(t))$ on some
Markovian stochastic variable $y(t)$ whose  PDF $\eta(y,t)$
satisfies equation
\begin{equation} \label{nmark3}
\dot \eta = -\hat {\Lambda}\eta, \;\;\mbox{with} \;\;
\eta(y|0)=\eta_i(y).
\end{equation}
In this equation $\hat \Lambda$ is a linear operator in
$\{y\}$-space and $\eta_i (y)$ is the initial condition
[$\int\!dy\,\eta_i (y) = 1$].

The dependence $\hat {\cal L}_1 (y)$ is taken in the form $\hat
{\cal L}_1(y(t)) \equiv \delta({y_0\!- \!y(t)})\hat {\cal L}_1 $,
where $y_0$ is the coordinate at which the system undergoes the jump
described by $\hat {\cal L}_1 $. Similar to the above-considered
model of $z(t)$-modulation of $\hat {\cal L}_1$, in the case of
$y(t)$-modulation the evolution of the system is described by the
PDF $p(x,y|t)$ in the combined space $ \{x \otimes y\}$. This PDF
satisfies the SLE of type of eq. (\ref{field9}), which as applied to
the Laplace transform $P (\epsilon) = \int_0^{\infty} \!dt \, p
(t)e^{-\epsilon t}$ is given by
\begin{equation} \label{nmark4}
\epsilon P = \delta({x\!-\!x_i})\delta({y\!-\!y_i}) -[\hat
{\Lambda}+\delta({y\!-\!y_0})\hat {\cal L}_1]P.
\end{equation}
Of special interest is the PDF averaged over $y(t)$-process
\begin{equation} \label{nmark5}
\overline{\rho}_y (x,x_i|t) = \!\int \!dy\!\int \!dy_i \,
P(x,y;x_i,y_i|t) \eta_i(y_i).
\end{equation}

The $y(t)$-controlled (or modulated) process in $\{x\}$-space proves
to be of CTRW type. Thus obtained CTRW depends on the initial
condition $p_i (y)$ and the form of the operator $\hat \Lambda$. In
what follows we will consider the non-stationary variant realized
for $p_i (y)=\delta({y\!-\!y_0})$ \cite{Shu1}. In this variant the
average PDF $\overline{\rho}_y (x,x_i|t)$ is known to satisfy the
CTRW-like equation usually written as applied to $R(\epsilon) =
\overline{R}_y (\epsilon) = \int_0^{\infty} \!dt \,
\overline{\rho}_y (t) e^{-\epsilon t}$ \cite{Shu1}:
\begin{equation} \label{nmark6}
\epsilon R (\epsilon) = \rho_i - {M}(\epsilon) \hat {\cal
L}_{\alpha} R (\epsilon),
\end{equation}
in which
\begin{equation} \label{nmark7}
 M(\epsilon) = \overline{M}(\epsilon) = (D_1/D_{\alpha})\epsilon
 \langle y_0 |(\epsilon + \hat \Lambda)^{-1}|y_0
 \rangle.
\end{equation}

The behavior of $\overline{M}(\epsilon)$ is completely determined by
the specific features of the controlling process (\ref{nmark3}).
Various models of this process are discussed in ref. \cite{Shu1}.

\section{CTRW-based SLE.}

The Markovian representation is very suitable for treating the
effect of time-dependent field on CTRW-like processes.

The corresponding equation is straightforwardly derived by taking
into account that, in accordance with the Markovian representation,
this equation describes Markovian process in $\{x \otimes y\}$-space
affected by the the driving force which can be modeled by
interaction with the Markovian $z(t)$ variable. This means that the
equation sought is actually the Markovian SLE \cite{Kubo} for the
PDF $q({\bf r};{\bf r}_i|t)$ in the extended space $\{{\bf r}\}
\equiv \{x \otimes y \otimes {\bf z}\}$-space. For the Laplace
transform $Q (\epsilon)= \int_0^{\infty} \!dt \, q(t) e^{-\epsilon
 t}$ this equation is written as:
\begin{equation} \label{sle1}
\epsilon Q = \delta_{{\bf r}{\bf r}_i} - (\hat
{\Lambda}+\delta_{yy_0}\hat {\cal L}_1 + \hat L) Q
\end{equation}

This equation is seen to differ from eq. (\ref{nmark4}) for
$P(x,y|\epsilon)$ only in the replacement $\epsilon$ by $\hat \Omega
= \epsilon + \hat L$ and the evident change of $\delta$-function
describing the initial condition. Naturally, for the PDF averaged
over $y(t)$-process:
\begin{equation} \label{sle2}
R(\epsilon) = \overline{R}_y (\epsilon) = \int_0^{\infty} \!dt \,
\overline{\rho}_y(t) e^{-\epsilon  t},
\end{equation}
one gets CTRW-like equation (sometimes called the non-Markovian SLE
\cite{Shu1}) similar to eq. (\ref{nmark6})
\begin{equation} \label{sle3}
\big[\hat \Omega + \hat {\cal L}_{\alpha}{M} (\hat \Omega)\big]R =
\rho_i\sigma_i   \;\;\mbox{with}\;\;\hat \Omega = \epsilon + \hat L.
\end{equation}
Notice that the order of operators $\hat {\cal L}_{\alpha}$ and ${M}
(\hat \Omega)$ is important since they do not commute with each
other.

To qualitatively interpret eq. (\ref{sle3}) within the CTRW approach
it is worth noting that, according to the SLE representation
(\ref{field9}), the time-dependent-field affected CTRW can be
considered as sequence of jumps governed by the $z$-dependent
operator  ${\cal L}_1(z)$. The CTRW process is accompanied by the
simultaneous evolution of the parameter $z(t)$, determined by the
operator $e^{-\hat Lt}$. This operator will enter in formulas
describing CTRW evolution in the form of the product $W(t)e^{-\hat
Lt}$ which for the Laplace transform $\widetilde{W}(\epsilon)$ just
corresponds to the replacement $\epsilon$ by $\hat \Omega = \epsilon
+ \hat L$ in $\widetilde{W}(\epsilon)$ thus resulting in eq.
(\ref{sle3}).

Exact eq. (\ref{sle3}) essentially differs from that proposed
earlier \cite{Sok,Han1} to treat the effect of time-dependent field,
i.e. the results of these works are, in general, incorrect, except,
may be, some special cases (see below).

\subsection{General results.}

In this brief communication we will restrict ourselves to analyzing
the time evolution of the moments of the PDF
$\overline{\rho}_{yz}(x|t)$ (averaged over $y(t)$- and
$z(t)$-processes) $m_n(t) = \int \!dx \, x^n \overline{\rho}_{yz}(t)
$. For this analysis we need to specify the initial PDF $\rho_i
(x)$. For simplicity we will assume $\rho_i (x) = \delta (x)$.

Instead of moments $m_n(t)$, it is more convenient to analyze their
Laplace transforms $\widetilde{m}_n (\epsilon)$ which can be
expressed in terms of moment operators in the $\{{\bf z}\}$-space
\begin{equation} \label{app2}
\mbox{$\hat M_n = \int \!dx \, x^n {\overline{R}}_{y}(x,\hat
\Omega):$}
\end{equation}
\begin{equation} \label{app3}
\mbox{$ \widetilde{m}_n(\epsilon) = \langle \hat M_n\rangle_z = \int
\!dz \!\int \!d\bar z\, \langle z|\hat M_n|\bar z\rangle
\sigma_i(\bar z).$}
\end{equation}
As is seen from eq. (\ref{sle3}) the operators $\hat M_n(\epsilon)$
satisfy simple equations
\begin{equation}\label{app4}
\hat\Omega \hat M_n = nfzM(\hat \Omega)\hat M_{n-1} +
n(n-1)D_{\alpha}\hat M_{n-2}
\end{equation}
for $n\geq 2$ with $\hat M_{-1}=0$ and $\hat M_0 = \hat
\Omega^{-1}$, in which $f = D_{\alpha}F_0z_0$. Solution of these
equations and substitution into eq. (\ref{app3}) yields for Laplace
transforms of time derivatives of the moments
\begin{equation}\label{app5}
\widetilde{\dot m}_1 (\epsilon) = f \langle z \Phi(\hat
\Omega)\rangle_z,\;\; \widetilde{\dot m}_2(\epsilon) =
\widetilde{\dot \mu}_{0}(\epsilon) + \widetilde{\dot
\mu}_2(\epsilon),
\end{equation}
with
\begin{equation}\label{app6}
\widetilde{\dot \mu}_0(\epsilon) = 2D_{\alpha}\langle \Phi(\hat
\Omega)\rangle_z, \; \widetilde{\dot \mu}_2(\epsilon) =  f^2 \langle
(z \Phi(\hat \Omega))^2\rangle_z.
\end{equation}
Here $\Phi (\hat \Omega)  = \hat \Omega^{-1} M(\hat \Omega)$.

In what follows we will concentrate on the analysis of force
dependent terms. The force independent contribution $\widetilde{\dot
\mu}_0(\epsilon)$ was discussed in detail earlier \cite{Met1}.

After the inverse Laplace transformation of expressions (\ref{app5})
and (\ref{app6}) one gets $\dot \mu_0(t) = 2D_{\alpha}\phi(t)$,
\begin{eqnarray}
\dot m_1 (t) &=& f \mbox{$\int \!\!\!\int \!d{\bf z}d{\bf z}_i\,z
g({\bf
z},{\bf z}_i|t)$}\sigma_i({\bf z}_i), \label{app7a}\\
\dot \mu_2(t) &=&  \mbox{$f^2 \!\!\int \!\!\!\int\!\!\!\int \!d{\bf
z}d\bar{\bf z}d{\bf z}_i \,z \bar{z}\nonumber$}\\
&&\mbox{$\;\;\;\times\int_0^t \!d\bar t \, g ({\bf z},\bar{\bf
z}|t-\bar t) g(\bar {\bf z},{\bf z}_i|\bar t)\sigma_i({\bf
z}_i)$}.\;\;\;\; \label{app7c}
\end{eqnarray}
In these formulas
\begin{equation}\label{app8}
g(\bar {\bf z},\bar {\bf z}|t) = \phi (t) \langle {\bf z}| e^{-\hat
Lt}|\bar {\bf z} \rangle,
\end{equation}
where $\langle {\bf z}| e^{-\hat Lt}|\bar {\bf z} \rangle$ is
controlled by the model of $z (t)$-modulation, while $\phi (t) =
(2\pi i)^{-1} \int_{-i\infty}^{i\infty}\! d\epsilon\,
\epsilon^{-1}M(\epsilon)$ is determined by the CTRW model considered
and for the subdiffusion model [$M(\epsilon) =
\epsilon^{1-\alpha},\: (\alpha < 1)$]
\begin{equation}\label{app9}
\phi (t) = \Gamma^{-1} (\alpha)t^{\alpha - 1}.
\end{equation}

\subsection{Applications}

\paragraph{Harmonically oscillating force.}

In the model of harmonically oscillating force (\ref{field10}) in
which $\hat L$ [eq. (\ref{field11})] describes dynamical motion in
the harmonic potential, one gets $\langle {\bf z}| e^{-\hat Lt}|\bar
{\bf z} \rangle = \delta ({\bf z}-{\bf z}_c(\bar {\bf z}|t))$, where
${\bf z}_c(\bar {\bf z}|t) = ({ z}_c(\bar {\bf z}|t),{v}_{z_c}(\bar
{\bf z}|t))$ is the trajectory of dynamical motion with ${\bf z}_c
(t=0)= \bar {\bf z}$ in the phase space $\{{\bf z}\}$.

Substituting this formula into eqs. (\ref{app7a})-(\ref{app8}) one
obtains $\mu_0(t) = 2D_{\alpha}\Gamma^{-1}(1+\alpha)t^{\alpha},$
\begin{eqnarray}
m_1 (t) &=& \mbox{$f \int_0^t d\tau z_c(\tau)\phi(\tau)$},
\;\label{harm2a}\\
\mu_2(t) &=& \mbox{$f^2 \int_0^t \!d\bar t \, z_c(\bar
t)\int_0^{\bar t}\!d\tau\, z_c(\tau) \phi(\bar
t-\tau)\phi(\tau)$},\;\;\label{harm2c}
\end{eqnarray}
where $z_c(t)$ is given by eq. (\ref{field10}).

For brevity, we will restrict ourselves to the discussion of the
long time behavior of the moments only:
\begin{eqnarray} 
m_1 (t) &\stackrel{t \to \infty}{\simeq}& (f/\omega^{\alpha})
\sin(\varphi + \pi \alpha/2)\; \label{harm3a}\\
\mu_2(t) &\stackrel{t \to \infty}{\simeq}& \gamma_2 (\alpha)
f^2(t/\omega)^{\alpha}\;\;\label{harm3b}
\end{eqnarray}
where $f = D_{\alpha}F_0z_0$ and $\gamma_2 (\alpha) =
\cos(\pi\alpha/2)/[2\Gamma(1+\alpha)]$.

These formulas predict some peculiarities of the subdiffusion
response to oscillating force. First, $m_1 (t)$ appears to be
nonzero with asymptotic value (at $t \to \infty$) independent of
time and harmonically oscillating as a function of the initial phase
$\varphi$ of force oscillations. Second, the force dependent part of
$\mu_2 (t)$ is anomalously large increasing in time so that $\mu_2
(t)/\mu_0 (t) = \cos(\pi\alpha/2)
[f^2/(4D_{\alpha}\omega^{\alpha})]$ independent of time. Third, in
the case of conventional diffusion, i.e. at $\alpha \to 1$,
$\mu_2(t)/t^{\alpha} \to 0$, as should be.

The exact formulas (\ref{harm3a}) and (\ref{harm3b}) significantly
differ in their analytical form from those derived earlier with not
quite correct kinetic equation \cite{Sok}, though, surprisingly, the
results obtained with this equation appeared to be qualitatively
correct.

\paragraph{Stepwise oscillating force.}

Here we will briefly discuss the model of stepwise oscillating force
to check the results obtained in ref. \cite{Han1} with the equation
which is not quite correct, in general. The model is defined as
$z(t) = z_0 (-1)^{[2t/\tau_0]}$, where $\tau_0$ is the oscillation
period and $[x]$ denotes the integer part of $x$. It can also be
represented as
\begin{equation}\label{stp0}
z(t) = z_c(t) = \sum_{n=-\infty}^{\infty} z_ne^{i n \omega t},
\end{equation}
where $\omega = 2\pi/\tau_0$ and $z_{n}$ are given by: $z_{2n} =0,
\, z_{2n+1}=-2i/[\pi(2n+1)]$ with $z_{-n} = z_n^{*}$.

As mentioned above, the case of $z_c (t)$ represented as a
superposition of harmonically oscillating functions can be described
by assuming $z_c (t)$-modulation to result from dynamical motion in
a harmonic potential in the multidimensional space ${\bf z} =
(z_1,v_{z_1};\dots;z_n,v_{z_n};\dots)$. In this case formulas
(\ref{app7a})-(\ref{app7c}) predict the same expressions
(\ref{harm2a})-(\ref{harm2c}) for the moments. Evaluation with these
expressions yields:  $m_1 (t) \stackrel{t \to \infty}{\simeq} \bar
\gamma_1 (\alpha)(2f/\omega^{\alpha}),\;$ and
\begin{equation}\label{stp1}
\mu_2(t) \stackrel{t \to \infty}{\simeq} \bar \gamma_2 (\alpha)f^2
(t/\omega)^{\alpha}.\;
\end{equation}
Here the functions $\bar \gamma_1 (\alpha)$ and $\bar \gamma_2
(\alpha)$ are defined as:
\begin{eqnarray}\label{stp2}
\bar \gamma_1 (\alpha)& =& \zeta(1+\alpha) (2-2^{-\alpha}) \sin
(\pi\alpha/2),\\
\bar\gamma_2 (\alpha) &=& \zeta (2+\alpha)
(4-2^{-\alpha})/[4\Gamma(1+\alpha)],
\end{eqnarray}
where $\zeta (x)$ is the Riemann's zeta-function.

Obtained results agree with those of ref. \cite{Han1} thus
confirming the correctness of the method proposed in this work as
applied to the model of stepwise oscillating force.

\paragraph{Fluctuating force.}

Of great interest is also the case of fluctuating force $F(t)$, i.e.
fluctuating $z(t)$. This case is conveniently analyzed using eqs.
(\ref{app5}) and (\ref{app6}) for the Laplace transforms.

For brevity, we will analyze only the long time limit. In addition,
for simplicity, we will assume that the initial condition $\sigma_i
(z)$ in $\{{\bf z}\}$-space is equilibrium i.e. $\hat L \sigma_i =
0$, and $m_1 (0)\equiv\langle z \rangle_{\sigma_i} = 0$. In such a
case the first moment $m_1(t) = m_1(0) = 0$ so that the only value
to be analyzed is $\widetilde{\mu}_2 (\epsilon)$. At small $\epsilon
\to 0$ this term can be estimated as
\begin{equation}\label{flu1}
\widetilde{\mu}_2 (\epsilon) \approx f^2
\epsilon^{-1}\Phi(\epsilon)\langle e_z| z \Phi (\hat \Omega)
z)|e_z\rangle,
\end{equation}
where $|e_z \rangle$ and $\langle e_z|$ are the equilibrium states
of the operator $\hat L$ (in this bra-ket notation $\langle e_z|
\equiv \int\! d{\bf z}$ \cite{Shu1} and $| \sigma_i \rangle = | e_z
\rangle $) therefore
\begin{equation}\label{flu2}
\mu_2 (t) \stackrel{t \to \infty}{\simeq} \Gamma^{-1}(1+\alpha)\bar
f^2 (\bar\tau t)^{\alpha}
\end{equation}
with $\bar f = D_{\alpha}F_0 \sqrt{\langle z^2\rangle}$ and the
parameter $\bar \tau$ is expressed in terms of the correlation
function $K(t) = \langle z z(t) \rangle = \langle e_z| z e^{-\hat L
t}z|e_z\rangle$:
\begin{equation}\label{flu3}
\bar \tau^{\alpha} = \int_0^{\infty}\!dt\, K(t)\phi(t)/\langle
z^2\rangle.
\end{equation}

It is seen that the characteristic features of $\mu_2 (t)$ are
similar to those obtained in the case of oscillating force with $f$
and $\omega$ replaced by $\bar f$ and $\bar\tau^{-1}$, respectively.
Noteworthy is, however, that unlike this case, for fluctuating force
$\mu_2 (t)/t^{\alpha}$ is finite as $\alpha \to 1$.

\section{Concluding remarks.}

We have analyzed the response of CTRWs on time-dependent field using
the rigorous method  based on the Markovian representations of CTRW
and the modulated field. This method is applied to describing the
field effect on subdiffusive motion.

Obtained formulas (\ref{harm3a})-(\ref{flu2}) clearly demonstrate
some specific features of the response of anomalous subdiffusive
systems: 1) in the case of oscillatory force modulation the first
moment (average displacement) is, in general, non-zero (even in the
long time limit) and depends on the oscillation phase, 2) the
modulated force results in the anomalously strong contribution to
the second moment (dispersion) growing in time.

In conclusion, it is worth noting that the proposed Markovian SLE
approach (\ref{field9}) for describing the influence of modulated
external fields is applied to only one particular problem of the
theory of force induced effects in stochastic systems. This approach
is, however, fairly general and can be very suitable in studying
many time-dependent-field affected stochastic processes
\cite{Han,Rei,Dyk} since it reduces the study to the analysis of
characteristic features of time-independent operators (their
spectra, eigenfunctions, etc.). Some of applications of the SLE
approach (\ref{field9}) are currently under consideration.

\textbf{Acknowledgements.}\,  This work was partially supported by
the Russian Foundation for Basic Research.

\end{document}